\documentclass[aps,prl,twocolumn,superscriptaddress,showpacs]{revtex4-1}

\usepackage{amsfonts,amssymb,amsmath,latexsym,epsfig,wasysym,bbold} 
\usepackage[sort&compress]{natbib}
\usepackage{float}
\usepackage{placeins}
\usepackage{rotating}
\usepackage[normalem]{ulem}

\begin{document}

\title{Sampling  motif-constrained ensembles of networks}

\author{Rico Fischer}
\affiliation{Max Planck Institute for the Physics of Complex Systems, 01187 Dresden, Germany}
\author{Jorge C. Leit\~ao}
\affiliation{Max Planck Institute for the Physics of Complex Systems, 01187 Dresden, Germany}
\author{Tiago P. Peixoto}
\affiliation{Institut f\"ur Theoretische Physik, Universit\"at Bremen, Hochschulring 18,
  28359, Bremen, Germany}
\author{Eduardo G. Altmann}
\affiliation{Max Planck Institute for the Physics of Complex Systems, 01187 Dresden, Germany}

\begin{abstract}
  The statistical significance of network properties is 
  conditioned on null models which satisfy specified properties but 
  that are otherwise random.  Exponential random graph models are a
  principled theoretical framework to generate such constrained ensembles, but which
  often fail in practice, either due to model inconsistency, or due to the
   impossibility to sample networks from them. These problems
  affect the important case of networks with prescribed clustering
  coefficient or number of small connected subgraphs (motifs).  In this Letter we use the Wang-Landau method to
  obtain a multicanonical sampling that overcomes both
  these problems.  We sample, in polynomial time, networks with arbitrary degree
    sequences from ensembles with imposed motifs counts.  Applying this method
    to social networks, we  investigate the relation between transitivity and homophily,
    and we quantify the correlation between different types of motifs, finding that
    single motifs can explain up to $60\%$ of the variation of motif profiles.
 \end{abstract}


\maketitle

Networks form the basis of an ample class of complex systems. The
observed topological patterns of such systems often yield the only
available evidence for the underlying principles behind their
formation. However, the significance of any observed property can only
be assessed in comparison to a properly defined network ensemble that acts as a ``null''
model~\cite{NunesAmaral2006,Holme2007,Newman2010}.  For instance,
clustering (i.e. high density of triangles), skewed degree distributions, and community structure are considered significant in real networks
because they are absent in Erd\H{o}s-Renyi networks. To perform such
comparisons, it is essential not only to properly define such null
models, but also to correctly sample network realizations from
them. This is relatively straightforward when the ensemble
generates networks where the edges are sampled independently
(e.g. Erd\H{o}s-Renyi and configuration
models~\cite{Newman2001a,Chung2002}, the stochastic
block model~\cite{Holland1983, Karrer2011}) and it remains feasible when
strict edge independence is violated due to hard constraints~\cite{Blitzstein2010,Kim2014,Orsini2015}. 
 However,  for ensembles with more generic constraints the sampling is
  significantly more challenging. A 
particularly important example is ensembles with a prescribed density
of connected subgraphs
(``motifs'')~\cite{Strauss1975,Park2004a,Foster2010}. For
this class of models, one often finds abrupt phase transitions, where
sampled networks possess either very high or very low motif
density~\cite{Foster2010, Park2004a}, excluding intermediary
values often encountered in real systems. Furthermore, they often show
strong non-ergodic behavior, with very slow relaxation that forbids
unbiased sampling in practical computational time~\cite{Foster2010}. Since the edge
placement is not independent, the densities of different motifs are
correlated with each other and also with
large-scale network structures~\cite{Foster2011,Beber2012}. Without addressing the
issue of correct sampling, these correlations cannot be
properly identified, which makes the occurrence of these patterns in
real systems difficult to interpret. In particular, it is not possible to
conclude whether a particular motif density profile indicates a topology
optimized towards robustness~\cite{Milo2002,Milo2004} or whether it is
merely a byproduct of a specific large-scale
structure~\cite{Foster2011,Artzy-Randrup2004}, of combinatorial constraints
  \cite{Ugander2013}, or of correlations between motifs.

In this Letter we show how to sample from ensembles with
prescribed motif densities in polynomial time. We employ a
multicanonical Monte Carlo method~\cite{Landau2013} that allows the entire
range of the order parameter to be explored. In this manner, not
only the non-ergodicity problem is explicitly avoided, but it also 
becomes possible to sample networks with arbitrary motif densities, even
those at intermediate values that are unattainable via traditional
importance sampling.  This allows us to quantitatively investigate two  fundamental 
problems in social networks: the homophily-transitivity relationship and the interdependence of
different motif types.

\begin{figure}[!t]
\includegraphics[width=1\columnwidth]{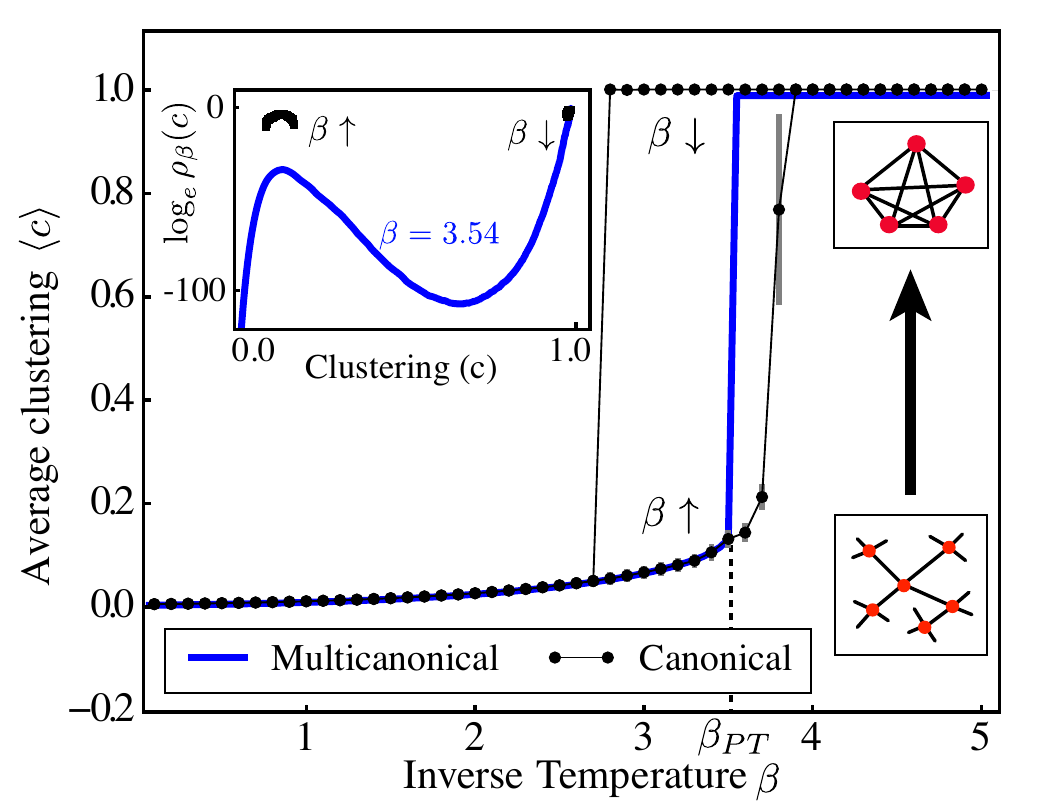}
\caption{ (Color online) Multicanonical sampling of exponential random graphs with
  imposed clustering avoids the limitations of canonical sampling.  The
  ensemble $\mathcal{G}$ corresponds to k-regular undirected networks
  with $N=640$ and degree $k=4$.  The observable is the clustering
  coefficient $s=c$ (proportional to the number of triangles,
  $n_\triangle$). The main plot shows $\langle c \rangle$ (and standard deviation) as a function of the inverse
  temperature $\beta$   obtained by canonical (symbols) and
  multicanonical (continuous thick line) sampling. Inset: the distribution
  $\rho_{\beta}(c)$ for $\beta = 3.54 \approx \beta_{PT}$ obtained by the two methods.
  Canonical samplings used $5\times 10^5$ MCMC steps 
  for equilibration, before another $5\times 10^5$ steps were used for
  estimation.  After these steps, the value of $\beta$ was slowly
  increased ($\beta\uparrow$) or decreased ($\beta\downarrow$) and the
  process repeated.  The multicanonical sampling used $20$ Wang-Landau
  steps to estimate $\rho_0(c)$ (each step used $5$ tunneling steps)~\cite{Dayal2004,Landau2013}. 
  }
\label{fig.1}
\end{figure}


We are interested in network ensembles that
possess one particular observable $s$ of interest, but that are
otherwise maximally random. The last requirement is essential to ensure
  that the ensemble is representative of the networks with a given $s$ and is
  not subject to additional (hidden) constraints. Both features are achieved by sampling
the network from an exponential random graph model
(ERGM)~\cite{Strauss1986,Snijders2002,Park2004,Newman2010,Horvat2015,Orsini2015} $\mathcal{G}$,
where each graph $g\in \mathcal{G}$ occurs with probability
\begin{equation}\label{eq.Pi}
\Pi_\beta(g) = \frac{e^{\beta s(g)}}{Z_\beta},  \text{ where } Z_\beta= \sum_{g\in\mathcal{G}}
e^{\beta s(g)},
\end{equation}
where $s(g)$ is the observable associated with network $g$, and $\beta$
is an inverse-temperature parameter, in analogy to the canonical
ensemble in statistical physics. The distribution of $s$
is $\rho_\beta(s) = \sum_{g\in \mathcal{G}} \delta(s(g)-s) \Pi_\beta(g)=
\rho_0(s) e^{\beta s}/Z_\beta$, where $\rho_0(s)\equiv \rho_{\beta=0}(s)$ is
called the state density (the fraction of networks~$g$ in the ensemble that
have observable equal to $s$). The ensemble that acts as a null model for an empirical network with
  $s=s^*$ is usually constructed fixing $\beta$ in such a way that  $\langle s \rangle_{\beta} \equiv \sum_s s \rho_{\beta}(s)$ equals $s^*$.  The number of networks in this ensemble typically
grows exponentially with the number of nodes, and, thus, besides a small
set of observables $s$ that can be treated analytically, investigation
of ERGMs requires sampling networks $g$ from $\mathcal{G}$ using Monte
Carlo methods~\cite{Park2004}.

The usual approach of sampling from $\mathcal{G}$ is via Markov chain
Monte Carlo (MCMC) method works as follows: starting from one network $g\in\mathcal{G}$,
a new network $g' \in \mathcal{G}_n$ is proposed by choosing two links
at random and exchanging one of the nodes of each link, which preserves
the degree-sequence of the network~\cite{Maslov2002}. The proposed
network is accepted with the Metropolis-Hastings probability $A(g\mapsto
g') =\min\{1, e^{\beta(s(g')-s(g))}\}$ and the process is repeated from
$g'$ ($g$) if the proposal is accepted
(rejected)~\cite{NewmanBarkemaBook}. Since
the moves fulfill ergodicity and detailed balance, for sufficiently long times the values of $s$ in the
sampled networks $g$ are distributed as $\rho_\beta(s)$.
However, despite this asymptotic guarantee, in practice this method
often fails because the time to approximate $\rho_\beta(s)$ grows exponentially with the
number of nodes 
$N$. This happens whenever $\rho_\beta$ possesses more than one local 
maximum (minimum of the free energy) and the barriers between them grow with $N$.  As we
show below, this generically happens when the observables $s$ are related to motifs.

As an alternative to the \emph{canonical} (simple Metropolis) sampling method described above, we propose
a \emph{multicanonical} sampling to overcome the aforementioned
problem. This method aims to sample networks uniformly on a pre-defined observable range
$[s_{min},s_{max}]$,  thus overcoming the minima of $\rho_\beta(s)$  that are responsible for
  the weak performance of the canonical method. This is done by sampling the states
according to auxiliary ensemble with probabilities $\Pi'(g)
\propto 1/\rho_0(s(g))$, achieved by simply changing the
acceptance to $A(g\mapsto g')=\min\{1,
\rho_0(s(g'))/\rho_0(s(g))\}$~\cite{Landau2013}. However, in order to
perform this sampling we need to know the state density $\rho_0(s)$.  In
order to estimate it, we use the Wang-Landau
algorithm~\cite{Wang2001,Landau2013}, which, in short, constructs an adaptive
histogram to approximate $\rho_0(s(g))$~\cite{Note1}. After
convergence, $\rho_\beta(s)$ is estimated for all $\beta$'s reweighting $\rho_0(s)$ through $\rho_\beta(s) =
\rho_0(s) \exp(-\beta s)/Z_\beta$~\cite{Landau2013}. Hence, the auxiliary ensemble allows to
explore the original canonical ensembles without being
restricted to the most probable regions.
More importantly, we can impose the desired value of the observable  as a hard
constraint a posteriori, i.e., only sample networks with $s(g)=s^*$. 
 The multicanonical approach has recently been
applied to investigate the spectral gap of networks~\cite{Saito2011},
and related approaches have been used to investigate
percolation~\cite{Hartmann2011a} and resilience
properties of networks~\cite{Dewenter2015}.

\begin{figure}[!bt]
\includegraphics[width=\columnwidth]{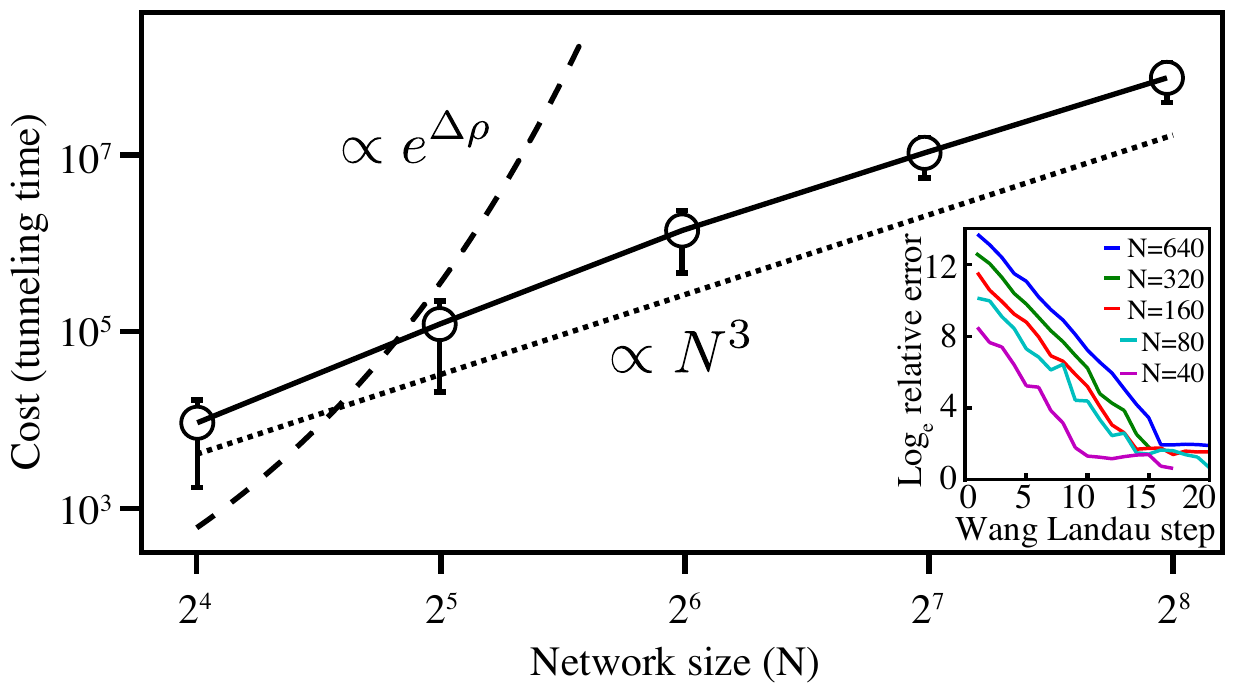}
\caption{ (Color online) Efficiency of the multicanonical method to sample networks with constraints.
 The computational cost (in number of MCMC steps) to generate an independent realization
 of a network in the $k$-regular ensemble with $k=3$ is plotted as a function of $N$.
  In the canonical method close to the critical $\beta$, this requires passing the minimum
  of $\rho_\beta(c)$ (inset of Fig.~\ref{fig.1}). We measured that the height of this barrier
  increases as $\Delta \rho \approx 0.4 N$, which leads to an
  exponential increase in the cost (dashed line).  Sampling independent realizations in
  the multicanonical method requires, at most, a tunneling (the number of MCMC steps to do $c=0 \mapsto c = 1 \mapsto
  c=0$)~\cite{Dayal2004}. The measured tunneling time (circles and full line)
  scales polynomially.
  Inset: convergence of the relative error in the logarithm of the density of states (entropy) during convergence of the Wang-Landau algorithm,
  estimated comparing the measured value with the exact value on c=1. The saturation of
  the error observed for large number of steps does  not hinder the sampling of any $c$
  (see Ref.~\cite{Belardinelli2008a} for methods to overcome the saturation).
  }
\label{fig.2}
\end{figure}

In Fig.~\ref{fig.1} we show how the application of multicanonical
sampling solves the limitations of canonical sampling in the classical
problem of introducing clustering in a $k$-regular
network~\cite{Strauss1975,Foster2010}. Here, nodes are forced to have the same degree $k$
and the observable of interest is the number of triangles, $s(g) = n_{\triangle}$. Fixing
$n_{\triangle}$ is the same as fixing the clustering coefficient
$c=3n_{\triangle}/n_{\wedge}$, where $n_{\wedge}$ is the 
number of connected triples (a constant for all networks with
the same degree sequence)~\cite{Newman2010}. This model exhibits a
transition at a specific value of $\beta=\beta_{PT}$ ($\approx 3.54$ for $k=4$), separating low and
high-clustering phases~\cite{Foster2010}.  The canonical sampling is unable to compute $\langle c
\rangle$ close to the phase-transition because
it yields different estimations of $\langle c \rangle$, depending
whether $\beta$ is slowly increased ($\beta\uparrow$, lower branch) or
decreased ($\beta\downarrow$, upper branch).  This hysteresis is
typical around first-order phase transitions (coexisting phases) and indicates that the canonical sampling is in a metastable
state.  Indeed, $\rho_{\beta_{PT}}(c)$ has two local maxima in which the canonical
sampling becomes trapped (inset in Fig.~\ref{fig.1}). On the other hand, the multicanonical sampling
is immune to these problems: it correctly characterizes $\langle c
\rangle$ at $\beta=\beta_{PT}$ and reveals the full distribution
$\rho_{\beta=\beta_{PT}}$. Hence, the method is not only capable of
computing the correct ensemble average for any $\beta$, it yields
typical networks with any value of $c$, including the
significant gap $c\in [.2, 1]$ which is unattainable with the canonical
sampling.  In Fig.~\ref{fig.2} we confirm that the computational
cost of the multicanonical method 
scales polynomially with system size, a dramatic improvement over the exponential scaling
of the canonical method.

\begin{figure}[!bt]
\includegraphics[width=\columnwidth]{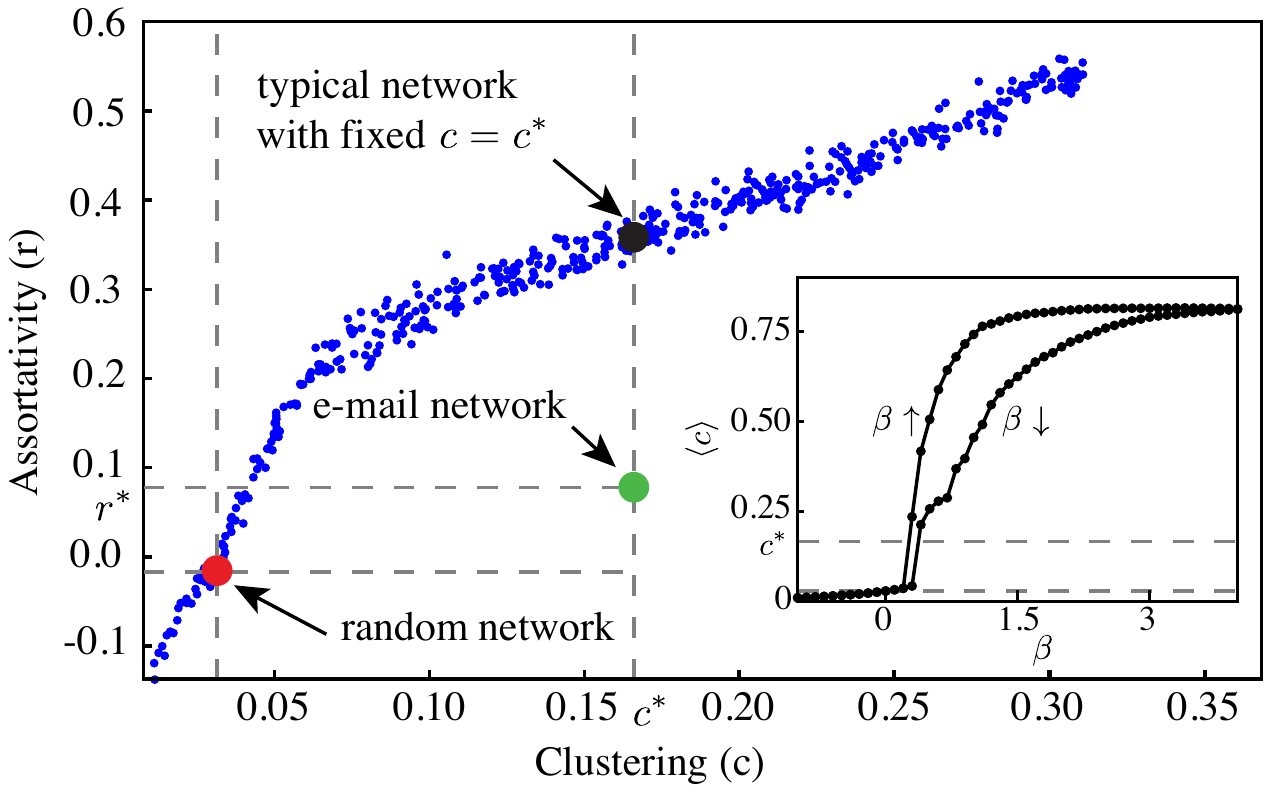}
\caption{ (Color online) Relationship between clustering~$c$ and assortativity~$r$ in the email
network of Ref.~\cite{Guimera2003}. The assortativity
$r^*$ is higher than in a random network with the same degree sequence,
but lower than in a typical network with a fixed
clustering $c^*$.  The plot shows $r$ and $c$ of different networks: the email network
(green), a typical fully random network (red), a typical random network with $c=c^*$
(black), and networks sampled using the multicanonical method from an ensemble with
equal probability for networks with the same $c$ (blue
dots). Inset: $\langle c \rangle$ obtained using the canonical method.}
\label{fig.3}
\end{figure}

\begin{figure*}[bt]
\includegraphics[width=1.8\columnwidth]{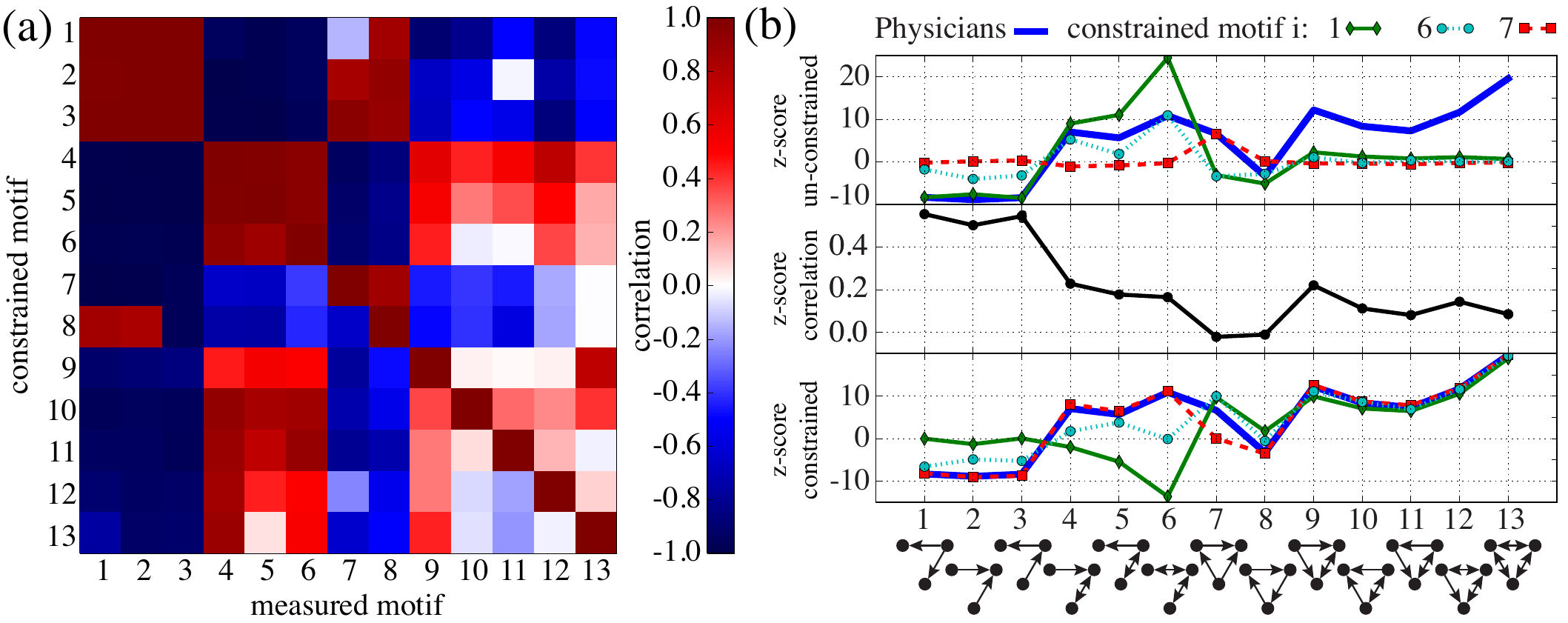}
\caption{ (Color online) Motifs are correlated to each other in blocks. (a) The Pearson correlation
  coefficient [$R_{ij} = (\langle n_in_j \rangle-\langle n_i \rangle
  \langle n_j \rangle)/\sigma_{n_i} \sigma_{n_j}$] between motifs $i$
  and $j$, computed by varying the constrained motif (in a range which
  includes the values of the real and random networks
seem~\cite{SM}
). (b) Upper panel:
  Motif profile~\cite{Milo2004} built from the z-score $z_j$ vs. $j$
  ($z_j = (n_j - \langle n_j \rangle) / \sigma_{n_j}$, where $n_j$ is
  the number of motifs $j$ and $\langle \ldots \rangle$ and $\sigma_j$
  are the average and standard deviation in the $\beta=0$ ensemble). Different lines
  correspond to the $z_j$ of the real network ($z_j^*$, blue line) and the expected
  $z_j$'s  in the constrained ensemble in which $n_i$ is equal to the
  $n^*_i$ of the real network, where $i$ is the constrained motif shown in the legend ($z_j=z_j^*$ for $j=i$). Middle
  panel: the  correlation between the profiles shown in the upper
  panel, i.e., between the profile $z$ of the real network and the
  profile $z'$ of the motif-$i$ constrained network (as a function of
  $i$), computed as $R_{zz'} = (\langle zz' \rangle -\langle z
  \rangle \langle z'\rangle)/\sigma^2_z$, where $\langle \ldots \rangle$
  and $\sigma_z$ were computed over $j\neq i$. Lower panel: comparison
  of the z-score shown in the upper panel (blue line) and the
  alternative z-score obtained computing $\langle \ldots \rangle$ and
  $\sigma_j$ in the ensemble constrained by $n_i=n_i^*$,  where $i$ is indicated in the legend
  ($z_j\equiv0$ for $j=i$).}
\label{fig.4}
\end{figure*}

Next we use the multicanonical method to investigate two important problems
of social networks.  The first problem we consider is to distinguish
between homophily (the tendency of ``similar'' nodes to connect to each
other) and transitivity (the tendency of nodes that already share a
common neighbor to connect to each other) in social
networks~\cite{Rapoport1953,Granovetter1973,Holme2007,Kossinets2009,Foster2011,Bianconi2014}. We
use the (undirected) network of email exchange within a
university~\cite{Guimera2003}.  It consists of $N=1,133$
users, and $M=5,451$ email exchanges, and a roughly exponential degree
distribution.  As observables we consider the clustering coefficient $c$
and the degree assortativity $r$~\cite{Newman2002}, for
which we obtain $c^*=0.166(12)$ and
$r^*=0.08 (3)$ (uncertainties in the last digit estimated using the order-10 Jackknife
method). We assess the significance of these values 
by comparing them to those obtained in the following three network
ensembles with the same degree sequence as in the original network: \\
\noindent (i) Same weight to all networks $g$ (i.e. the configuration
model). Canonical sampling with $\beta=0$ yields $\langle c
\rangle_{\beta=0} =0.028 (1)$ and $\langle r_{\beta=0}
\rangle =-0.017(13)$, much smaller than $c^*$ and $r^*$ as typically found in social networks.\\
\noindent (ii) ERGMs with $\langle c \rangle = c^*$. 
In order to
determine whether the assortativity is a consequence of high
clustering~\cite{Foster2011} we would like to measure $\langle r \rangle$ from
the null model with $\langle c \rangle = c^*$.  This canonical
sampling fails because  $\langle c \rangle_\beta$ vs. $\beta$ shows an
hysteresis around $s=c^*$ (inset of Fig.~\ref{fig.2}, in agreement with our previous discussion).\\
\noindent (iii) Hard constraints with $c(g)=c^*$, obtained using
multicanonical sampling. As mentioned before, this type of hard
constraint is unfeasible with canonical sampling, even if the desired
observable value is realizable. With the multicanonical method we sample
points after a number of Monte Carlo steps proportional to the tunneling
time, which guarantees that the sampled points are independent and
unbiased~\cite{Dayal2004}. We performed multicanonical sampling for a desired $c$ and
measured the assortativity $r$. The results are
shown in Fig.~\ref{fig.3} and reveal that random networks with the same
clustering of the email network $c=c^*$ typically show a much larger
assortativity $\langle r\rangle >r^*$. Therefore, although both $c^*$
and $r^*$ are larger than one would expect for a fully random network,
the actual value of $r^*$ is significantly less than one would expect by
knowing only $c^*$. From this we conclude that the degree homophily is not
  explained alone by transitivity.

The second problem we address is the extent to which the occurrence of
different motifs (connected subgraphs) are related to each other and the
impact of such correlations on the so-called motif
profiles~\cite{Milo2004}. Here we focus on directed networks, and the
observable of interest is the number $n_i$ of occurrences of a specific
motif $i$.  Again, traditional
sampling methods are not suited to address this problem because of the
existence of (potentially multiple~\cite{Foster2010}) discontinuous
phase transitions. Instead, using the multicanonical method, we 
reliably sample networks with a prescribed count of one particular
motif. By measuring the counts of all other motifs, we obtain the
correlations between them and the constrained motif. In this manner, we
obtain 
\cite{SM}
the interdependence between all 13 different 3-node motifs in a
directed acquaintance network between
physicians~\cite{Coleman1957} (with $N=241$ nodes and
$M=1,098$ edges).  The results in Fig.~\ref{fig.4} reveal strong
positive and negative correlations between pairs of motifs. Two blocks
of motifs can be identified ($1,2,3,7,8$ and $4,5,6$,
Fig.~\ref{fig.4}a). Motifs show positive correlations within their
blocks and are anti-correlated with motifs in the other blocks (the
motifs $9$ to $13$ show a mixed behavior). Given that one motif is over
(under) represented, one should expect also an over representation in
motifs positively (negatively) correlated with it.  As a consequence of
this correlation, we find that single motifs explain up to $60\%$ of the
variance of the motif profile across the other 12 motifs
(Fig.~\ref{fig.4}b, upper and middle panel). Furthermore, if the
constrained ensembles are used to compute alternative $z$-scores, we
find that the resulting motif profiles vary dramatically depending on
the constraint, with some motifs $j$ showing variations from $z_j \gg 0$
to $z_j \ll 0$ (Fig.~\ref{fig.4}b, lower panel). This sensitivity of the
motif profile $z_j$ shows that such profiles bring
limited insights on the over- or under-representation of individual
motifs in a network. In particular, since such non-trivial profiles as
those seen in Fig.~\ref{fig.4}b can be obtained by imposing the occurrence
of a single motif, it is questionable whether conclusions regarding the
underlying formation mechanisms can be reliably reached from
them~\cite{Milo2004, Artzy-Randrup2004}. Nevertheless, the null models
considered here represent a principled approach of assessing the
relative significance of motif occurrences that is more meaningful than the usual
comparison to fully random networks.

In summary, we have shown that multicanonical sampling 
allows for an improved network generation and for the investigation of
problems which were otherwise intractable.  In particular, we  characterize
ERGMs in cases where the usual canonical sampling fails and we  sample networks imposing
hard constraints, an alternative to a direct sampling of ERGMs even when
the usual algorithms are feasible. Our analysis of empirical networks
demonstrates that using the multicanonical sampling allows the
investigation of the interdependence between network properties. In
particular, we quantified the correlation between clustering and
assortativity, and between different motifs, as well as the extent to
which their significance profiles can be explained by single
motifs. This opens the possibility of investigating the correlation
between motifs as well as other local-scale properties and the
large-scale structure of networks~\cite{Foster2011}, such as
communities, core-peripheries and many others. The systematic
disentangling of these diverse features is a crucial and open problem in
the identification of fundamental models of network formation.

\emph{Acknowledgments} We thank J. M. V. P. Lopes for insightful
discussions.  This work was partially funded by the University of
Bremen, under the program ZF04, and FCT (Portugal), grant SFRH/BD/90050/2012.

\bibliographystyle{apsrev4-1}

%


\newpage

\begin{center}
\LARGE{Supplemental Material}
\end{center}

\subsection*{Wang-Landau algorithm to sample networks}

The sampling algorithms used in the paper perform random walks in the space of constrained
networks. In our case, this space is built by networks with a fixed number of nodes and a
fixed degree sequence (e.g. the degree sequence of the e-mail network).  Besides this
space, the observable we are interested in characterizing $s(g)$ (e.g. number of triangles of the
network $g$) and its range of interest $[s_{\min},s_{\max}]$ are also chosen a priori.

The Wang-Landau algorithm performs a random walk in the space of constrained
networks that aims to visit equally often any value of $s \in [s_{\min},s_{\max}]$.
The outputs of this algorithm are: 1. a numerical approximation of the state
density $\rho_0(s)$; and 2. a set of random networks $g$ such that their observables $s(g)$
are uniformly distributed in $[s_{\min},s_{\max}]$. 
Below we describe the main steps of the Wang-Landau algorithm.            

The following quantities have to be initialized and evolve in time:
\begin{itemize}
	\item a network $g$, initially set to be an arbitrary network with $s(g)\in [s_{\min},s_{\max}]$; $s=s(g)$ represents its observable.
	\item a histogram-like list $S(s)$, for $s \in [s_{\min},s_{\max}]$ (binned if $s$ is continuous), that represents a discrete approximation of $\log \rho(s)$; it is initialized for all $s$ to $S(s)=0$.
	\item the Wang-Landau refinement parameter~$f$, initialized at $f=1$ (the minimum value $f_{\min}$ is set a priori, e.g. $f_{\min} = 2^{-12}$). 
\end{itemize}
The algorithm evolves according to the following rules:
\begin{enumerate}
	\item Repeat until a pre-defined number (e.g., 10) of round-trips are achieved:
	\begin{enumerate}
		\item Randomly propose a new network $g'$ in the space of constrained networks, and compute $s' = s(g')$ (see how below);
		\item Update $g,s$ to $g = g', s = s'$ if $\log(r) < a(s',s) = S(s') -
                  S(s)$ where $r$ is a random number drawn from a uniform distributed in $[0,1]$;
		\item Update $S(s) = S(s) + f$.
	\end{enumerate}
	\item update $f$ to $f = f/2$, go to 1. if $f \geq f_{\min}$.
\end{enumerate}

After the convergence of the evolution $(f<f_{\min})$ described above, $S(s)$ is an approximation of $\log \rho(s)$ up to a normalization constant. Setting $f = 0$ at this point and repeating loop 1 generates networks $g$ such that $s(g)$ is uniformly distributed in $[s_{\min},s_{\max}]$ (multicanonical ensemble).

Notes:
\begin{itemize}
	\item[{\it i})] A round-trip is achieved in step 1. when, for the first time, the observable went from $s = s_{\min}$ to $s = s_{\max}$ and returned back.
	\item[{\it ii})] The network $g'$ is constructed by selecting two edges of the original
          network $g$ and exchanging one of the nodes of one edge by one of the nodes of
          the other edge. This guarantees that the degree sequence is preserved. For more
          complicated constraints on which this procedure is not feasible, one can reject a proposed $g'$ if it is not in the space of constrained networks.
	\item[{\it iiii})] In order to achieve a faster computation of $s'=s(g')$ in 1(a) it is useful
          to store which motifs any given edge belongs to.   
          Then, instead of re-computing all motifs of $g'$ to compute $s'=s(g')$, one can
          calculate $s'$ from the number of motifs destroyed and created when passing from
          $g$ to  $g'$ using the procedure described in the note [{\it ii})].
	\item[{\it iv})] A canonic ensemble with fixed $\beta$ is obtained by running loop 1.
         with $a(s',s) = -\beta (s' - s)$. 
\end{itemize}

For a more formal description of the Wang-Landau algorithm and for further details we refer to
Refs.~\cite{Wang2001,Landau2013}. In Ref.~\cite{Note1} we provide an implementation of the
algorithm described above for the case of  triangles in undirected networks.

\bibliographystyle{apsrev4-1}

\newpage

\begin{sidewaysfigure}
\vspace{9cm}

\includegraphics[width=1.0\columnwidth]{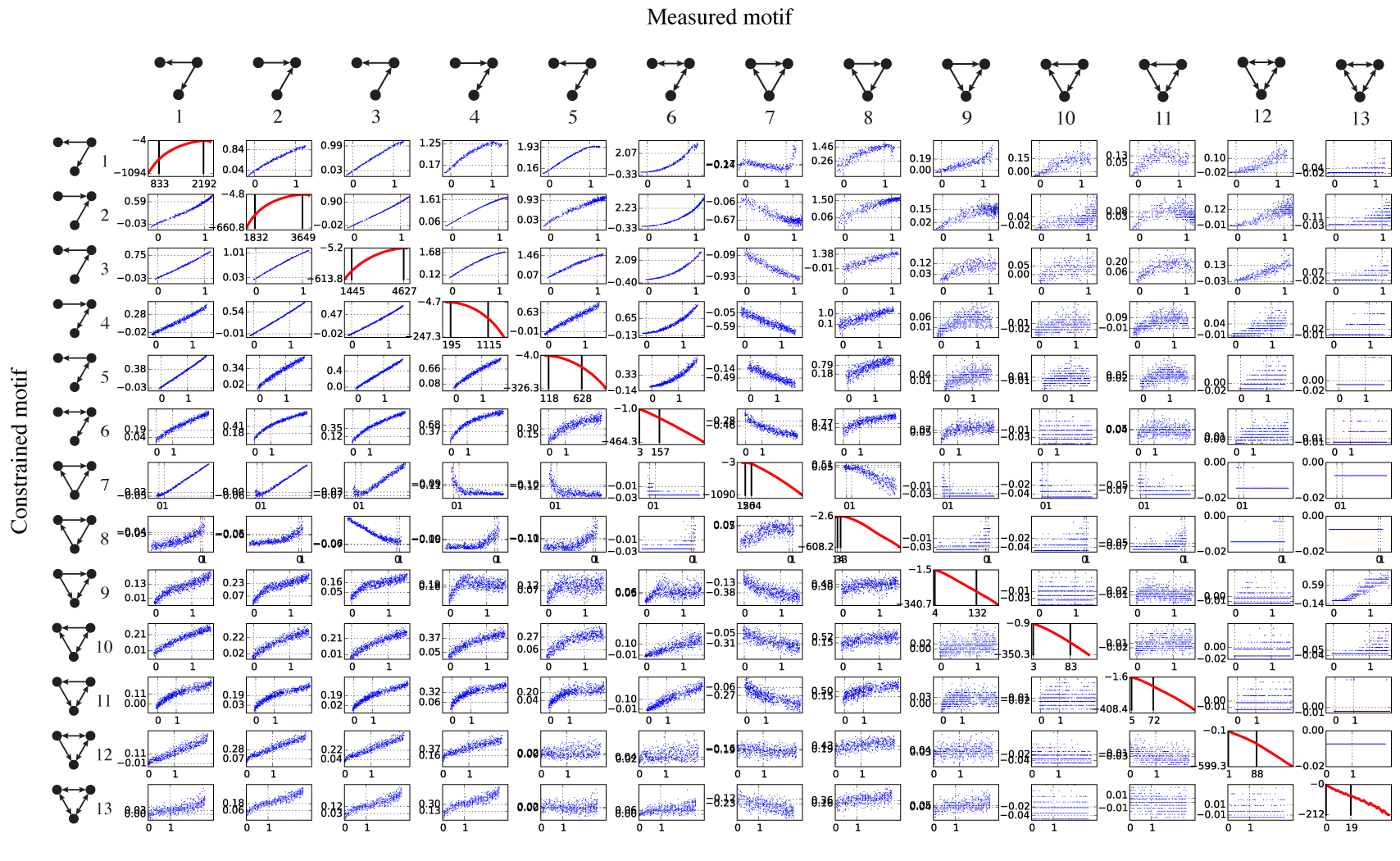}
\caption{ (Color online) Co-dependence between motifs in a directed social network (Physicians). The diagonal plots ($i=j$) show
  the probability of observing the motif $i$ in the ensemble of random networks with the
  same degree distribution (state density). Vertical lines indicate the expected number of motifs
  (close to the maximum) and the  number observed in the real network. The non-diagonal
  plots show the number of motifs $n_i$ and $n_j$ in random networks with fixed number of
  $n_i$ (constrained motif). For each constrained motif~$i$, a multicanonical 
sampling was performed in the range shown in the plot. For each value of the motif~$i$
(vertical axis) in this range, independent samples of networks were recorded and the number of
the unconstrained motifs $j$ (horizontal axis) was measured. The (Pearson) correlation between these
values is shown in Fig.~\ref{fig.4}(a). The motif-profile in Fig.~\ref{fig.4}(b) was
obtained over the values of motifs $j$ obtained fixing motif $i$ to the value of the real network.}
\label{fig.extra}
\end{sidewaysfigure}

\end{document}